\documentclass{article}

\usepackage{ifthen}
\newboolean{neurips_version}
\setboolean{neurips_version}{true}
\ifthenelse{\boolean{neurips_version}}
{\PassOptionsToPackage{numbers, compress}{natbib}
 \usepackage[preprint]{neurips_2022}%[final]
}
{\usepackage{arxiv}
 \usepackage[numbers]{natbib}
}

\usepackage[utf8]{inputenc} % allow utf-8 input
\usepackage[T1]{fontenc}    % use 8-bit T1 fonts
\usepackage{hyperref}       % hyperlinks
\usepackage{url}            % simple URL typesetting
\usepackage{booktabs}       % professional-quality tables
\usepackage{amsfonts,amsmath,amsthm,amssymb}       % blackboard math symbols
\usepackage{nicefrac}       % compact symbols for 1/2, etc.
\usepackage{microtype}      % microtypography
\usepackage{xcolor}         % colors
\usepackage[disable]{todonotes}
\usepackage{enumitem}
\usepackage{caption}
\usepackage{mathtools} %a labeled mapsto (specifically, \xmapsto from
\usepackage{tikz}
\usetikzlibrary{calc}
\usepackage{pgfplots}
\usepackage{graphicx}

\usepackage{amsmath}
\usepackage{algorithm, algpseudocode}
\usepackage{amsfonts}  % \mathbb{R}

\usepackage[english]{babel}
\usepackage{amsthm}

\usepackage{subfig}
\usepackage[export]{adjustbox}% in preamble
\usepackage{hhline}
\usepackage{makecell, caption, booktabs}
\usepackage{siunitx}

\usepackage{multirow}

\usepackage{amsmath, nccmath}
\usepackage{bigstrut}

\usepackage{booktabs}

\usepackage{lipsum}
\graphicspath{{media/}}     % organize your images and other figures under media/ folder

\usepackage{CJKutf8}
\usepackage[framed,numbered,autolinebreaks,useliterate]{mcode}
\title{Privacy-Preserving Logistic Regression Training on Large Datasets}

\author{%
\href{https://orcid.org/0000-0003-0378-0607}{\includegraphics[scale=0.06]{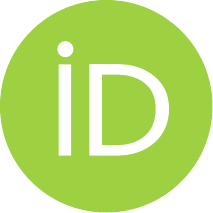}\hspace{1mm}John Chiang }
  \\
  \texttt{john.chiang.smith@gmail.com} \\
}
\date{}

\theoremstyle{remark}

\renewcommand{\epsilon}{\varepsilon}

\newcommand{\xdownarrow}[1]{%
  {\left\downarrow\vbox to #1{}\right.\kern-\nulldelimiterspace}
}
\newcommand{\xuparrow}[1]{%
  {\left\uparrow\vbox to #1{}\right.\kern-\nulldelimiterspace}
}

\makeatletter
\def\namedlabel#1#2{\begingroup
    #2%
    \def\@currentlabel{#2}%
    \phantomsection\label{#1}\endgroup
}
\makeatother

\algnewcommand{\LeftComment}[1]{\Statex \(\triangleright\) #1}
\algnewcommand{\LineCommentStep}[1]{\Statex \textbf{[Step #1]:} }
\makeatletter
\newlength{\trianglerightwidth}
\algnewcommand{\LineComment}[1]{\Statex \hskip\ALG@thistlm $\triangleright$ #1}
\algnewcommand{\LineCommentCont}[1]{\Statex \hskip\ALG@thistlm%
  \parbox[t]{\dimexpr\linewidth-\ALG@thistlm}{\hangindent=\trianglerightwidth \hangafter=1 \strut$\triangleright$ #1\strut}}
\algnewcommand{\LeftLineCommentCont}[1]{\Statex \hskip\ALG@thistlm%
  \parbox[t]{\dimexpr\linewidth-\ALG@thistlm}{\leftskip=\algorithmicindent \hangindent=\trianglerightwidth \hangafter=1 \strut$\triangleright$ #1\strut}}

\begin{document}

\maketitle

\begin{abstract}%
Privacy-preserving machine learning is one class of cryptographic methods that aim to analyze private and sensitive data while keeping privacy, such as homomorphic logistic regression training over large encrypted data. In this paper, we propose an efficient algorithm for logistic regression training on large encrypted data using Homomorphic Encryption (HE), which is the mini-batch version of recent methods using a faster gradient variant called $\texttt{quadratic gradient}$. It is claimed that $\texttt{quadratic gradient}$ can integrate curve information (Hessian matrix) into the gradient and therefore can effectively accelerate the first-order gradient (descent) algorithms. We also implement the full-batch version of their method when the encrypted dataset is so large that it has to be encrypted in the mini-batch manner. We compare our mini-batch algorithm with our full-batch implementation method on real financial data consisting of 422,108 samples with 200 freatures. %Our experiments show that  Nesterov's accelerated gradient (NAG)
Given the inefficiency of HEs, our results are inspiring and demonstrate that the logistic regression training on large encrypted dataset is of practical feasibility, marking a significant milestone in our understanding.

\iffalse
In both the training phase and prediction phases, the decryption key would only be accessed by the data owner, and thus the privacy of sensitive data is guaranteed while the encryption is secure. It has played an important role in various areas like finance, genomics and medical field full of sensitive private data. 

In this paper, we propose two efficient algorithms for privacy-preserving logistic regression training and evaluate our algorithms on real large datasets. Firstly, we extend the enhanced full-batch  Nesterov's accelerated gradient (NAG) with $\texttt{quadratic gradient}$  to the enhanced mini-batch version. We show that the exponential-decay learning rate can be adopted for the enhanced mini-batch version of NAG. Secondly, we evaluate the two enhanced mini-batch and full-batch versions and compare their differences over large datasets. 

Our experiments show that the enhanced full-batch NAG is more suitable than the enhanced mini-batch NAG. Given the inefficiency of HE techniques, our findings are particularly encouraging as they demonstrate the practical feasibility of logistic regression training on large encrypted datasets, marking a significant milestone in our understanding.
\fi
\end{abstract}

\listoftodos

\section{Introduction}
Suppose that there were several financial institutions each of which possesses a piece of private data of their customers. Although collecting all of their data together for training would lead to learning a better model, sharing sensitive data with other institiutions would be risky in terms of data security or even be illegal in many countires.  
Homomorphic Encryption (HE) is an encryption scheme that can be used to dress this dilemma. With HE, private data can be shared in an encrypted form to train machine learning algorihtms using various homomorphic operations. Compared to other approaches requiring additional assumptions and conditions, this HE-based approach is so flexible that the training computation can be delegated to even an untrusted third party since it need not decrypting the encrypted data, leakaging no sensitive information.

However, advanteages come with costs: even basic homomorphic operations on ciphertexts are several orders of magnitude slower than the corresponding operations in the clear setting, not to mention the ciphertext refreshing. In addition, factional number arithmetic operations on ciphertexts  are quite expensive in terms of managing the magnitude of plaintext. Fortunately, various state-of-the-art research works~\cite{gentry2012homomorphic, halevi2021bootstrapping, cheon2017homomorphic, van2010fully, brakerski2014efficient,  brakerski2011fully,  brakerski2012fully,  brakerski2014leveled, lopez2012fly,  coron2014scale,  ducas2015fhew} have contributed to migate these problemss.

There have been several studies on privacy-perserving machine learning, including logistic regression, mainly involving two types of approaches:
\paragraph{HE-based approaches} Graepel et al.~\cite{graepel2012ml} proposed two binary classifiers, linear means and Fisher's linear discriminant, using homomorphic evaluation operations. In 2018 and 2019, the iDASH competitions have helped to accelerate the research on homomorphic logistic regression training~\cite{chiang2022privacy,cheon2018ensemble,kim2018logistic,chen2018logistic, kim2018secure,IDASH2018bonte,aono2016scalable,IDASH2018gentry,IDASH2019kim}.  All these works either are on small-scall training datasets consisting of only hundreds of samples with dozens of features, or require the multiplication depth to be bounded, performing a limited number of iterations. Along with the increases of the size of datasets and the number of iterations, it is not clear whether or not their implementations are scalable.   Our mini-batch algorithm, however, is scalable in that it can perform a large number of iterations with linear time complexity. There have also been reported studies on homomorphic neural network inference~\cite{bos2014private,bost2014machine,       gilad2016cryptonets,   li2018privacy,juvekar2018gazelle,bourse2018fast, chiang2022volleyrevolver, hesamifard2017cryptodl,chillotti2021programmable,kim2018matrix,chabanne2017privacy}. However, the prediction stage is much simpler than the training stage, especially in terms of multiplication depth and computational amount, and thus training HE-based neural networks~\cite{chiang2023privacyCNN} is still a daunting task.

\paragraph{MPC-based approaches}
Nikolaenko et al.~\cite{nikolaenko2013privacy} presented an multi-party computation-based protocol to train linear regression model, namely a combination of linear homomorphic encryption and the Yao's garbled circuit construction~\cite{yao1986generate}. Mohassel and Zhang~\cite{mohassel2017secureml} improved this protocol and applied it to logistic regression and neural network training. However, MPC-based solutions not only result in large communication overhead increasing drastically with the increase of the number participants, but also require all of the participants to be online. An approach using the two-party computation was proposed~\cite{mohassel2017secureml} to mitigate this issue but needs an additional assumption that the two delegating servers involved do not collude. Note that the HE-based approach can admit a compromised server, requiring no such assumption.

The most related work to ours is the paper of Han et al~\cite{han2018efficient, han2019logistic}. For the first time, they combine the approximate HE and the approximate bootstrapping for privacy-preserving machine learning (training), namely the logistic regression training on encrypted data. And they also propose various novel optimization techniques, such as the ones for the bootstrapping operation, which parallelizes a bootstrapping operation by carefully designing the partition of a training dataset to split a ciphertext in order not to reconstruct the split ciphertxts, significantly reducing the parallellization overhead.

The specific contributions and novelty in our work are as follows:
\begin{enumerate}
    \item We present a mini-batch version of an enhanced NAG method accelerated by a faster gradient variant, using all the optimization techniques from~\cite{han2019logistic}. We demonstrate that our enhanced mini-batch NAG method has better performance than the original mini-batch NAG method in terms of convengence speed. 
    
    \item We propose the full-batch version of the same enhanced NAG method for large datasets based on the mini-batch database encoding method. We compare this enhanced full-batch version with the enhanced mini-batch version in terms of time comsumed.

\end{enumerate}

\section{Preliminaries}

We use square brackets ``$[\ ]$'' to signify the index of a vector or matrix element in the subsequent discussion. For instance, for a vector $\boldsymbol{v} \in \mathbb{R}^{(n)}$ and a matrix $M \in \mathbb{R}^{m \times n}$, $\boldsymbol{v}[i]$ or $\boldsymbol{v}_{[i]}$ denotes the $i$-th element of vector $\boldsymbol{v}$, and $M[i][j]$ or $M_{[i][j]}$ represents the $j$-th element in the $i$-th row of matrix $M$.

\subsection{Homomorphic Encryption}
Since Rivest et al.~\cite{rivest1978data} proposd the concept of homomorphic encryption that allows computation on encrypted data, there have been several proposals supporting only a single operation. For example, ElGamal Encryption scheme~\cite{elgamal1985public} and RSA can be used for the multiplication of ciphertexts while Okamoto-Uchiyama scheme~\cite{okamoto1998new} and Pailler encryption scheme~\cite{paillier1999public} allow the addition of ciphertexts without decryption. However, the longstanding open problem is a homomorphic encryption scheme to support both addition and multiplication operations. It is important to support both the two  operations because an arbitrary computation function can be approximated by polynomials consissting of addition and multiplication.

In 2009, Gentry~\cite{gentry2009fully} introduced the first secure HE using lattice-based cryptography: (1) He first  constructed a Somewhat Homomorphic Encryption (SHE) scheme that supports only a limited number of all kinds of operations without decryption. In this scheme, the message encrypted will not be recovered from the corresponding ciphertext after some operations mainly because the too large noise growing in the ciphertext will destroy the message in the encryption; and (2) To address this, he introduced a novel $\textit{bootstrapping}$ procedure that can refresh a ciphertext with large noise to produce another ciphertext encrypting the same messge but with small noise to allow more operations. With this bootstrapping technique, he demonstrated that an arbitrary function on encrypted values can be evaluated by his SHE without ever accessing the secret key. The resulting schemes are called Fully Homomorphic Encryption (FHE) schemes.

In subsequent years, extensive design and implementation work have been suggested, improving upon the early implementation by many orders of magnitude runtime performance, most of which are based on one of the following three hardness problems: Learning with Errors (LWE) problem~\cite{regev2009lattices}, Ring-LWE problem~\cite{lyubashevsky2010ideal} and Approximate GCD problme~\cite{van2010fully}.

FHE schemes are inherently computationally intensive, and managing the magnitude of plaintext is another crucial factor contributing to performance, as highlighted by Jaschke et al. \cite{jaschke2016accelerating}.  Fortunately, Cheon et al.~\cite{cheon2017homomorphic} proposed an approximate homomorphic encryption scheme (CKKS) involving a $\texttt{rescaling}$ procedure to effectively eliminate this technical bottleneck. Their open-source implementation, $\texttt{HEAAN}$, supports efficient floating point operations over encrytped data by rounding the encrypted message to discard its insignificant figures, as well as other common operations. For achieving better amortized time in homomorphic computation, $\texttt{HEAAN}$ supports a parallel technique (known as $\texttt{SIMD}$~\cite{SmartandVercauteren_SIMD}) to pack multiple numbers into a single polynomial, and provides rotation operations on plaintext slots.

Like~\cite{han2018efficient}, we use the approximate HE scheme, $\texttt{HEAAN}$, in our homomorphic logistic regression training algorithm. The underlying HE scheme in $\texttt{HEAAN}$ is comprehensively detailed in \cite{kim2018secure, han2018efficient}, and the foundational theory of abstract algebra relevant to it can be found in \cite{artin2011algebra}. Supposing that there are two vectors $m_1$ and $m_2$ of complex numbers to be encrypted into two ciphertext $\texttt{ct}_1$ and $\texttt{ct}_2$, respectively, some of its homomorphic operations applied in our work are briefly described as follows: 
\begin{enumerate}
  
  \item $\texttt{Enc}_{pk}$($m_1$): returns a new ciphertext $\texttt{ct}_1$ that encrypts the message $m_1$.
  
  \item $\texttt{Dec}_{sk}$($\texttt{ct}_1$): returns a vector $m_1$ of complex numbers that is encrypted by the message $\texttt{ct}_1$.
  
  \item \texttt{Add}($\texttt{ct}_1$, $\texttt{ct}_2$): returns a new ciphertext that encrypts the message $m_1 + m_2$.
  
   \item \texttt{cAdd}($\texttt{ct}_1$, $c$): returns a new ciphertext that encrypts the message$m_1 + c$, where $c$ is a complex number.
   
  \item \texttt{Mul}($\texttt{ct}_1$, $\texttt{ct}_2$): returns a new ciphertext that encrypts the message $m_1 \times m_2$.
  
  \item \texttt{cMul}($\texttt{ct}_1$, $c$): returns a new ciphertext that encrypts the message$m_1 \times c$, where $c$ is a complex number.
  
  \item \texttt{iMul}($\texttt{ct}_1$): returns a new ciphertext that encrypts the message$m_1 \times i$, where $i$  is the square root of $-1$.

  \item \texttt{bootstrap}($\texttt{ct}_1$): returns a new ciphertext that encrypts the message$- m_1$.

\end{enumerate}

\subsection{Database Encoding} \label{basic he operations} 
Assume that the training dataset, a matrix $Z$ of order $(n+1) \times f$, consists of $(n + 1)$ samples ${x_i}$ with $f$ features and $(n + 1)$ corresponding labels ${y_i}$, throughout this paper. Since a sample ${x_i}$ and its label ${y_i}$ always happen together in the training process, an efficient encoding method used in~\cite{kim2018logistic} is to integrate / incorprate labels into the samples, resulting in the matrix $Z$ as follows:
\begin{align*}
Z = 
  \begin{bmatrix}
    z_{[0][0]} & z_{[0][1]} & \ldots & z_{[0][f]} \\
    z_{[1][0]} & z_{[1][1]} & \ldots & z_{[1][f]} \\
    \vdots     & \vdots     & \ddots & \vdots     \\
    z_{[n][0]} & z_{[n][1]} & \ldots & z_{[n][f]}
  \end{bmatrix}
\end{align*}
where $z_{[i][j]} = y_{[i]} \cdot x_{[i][j]}$ for $0 \le i \le n$ and $0 \le i \le f$. Note that the original first column of matrix $Z$ is ones inserted afterwards[in advance] for the bias of LR model and hence now $z_{[i][0]} = y_{[i]} \times 1 = y_{[i]}$.
Kim et al.~\cite{kim2018logistic} proposed an efficient database-encoding method using the $\texttt{SIMD}$ technique, leveraging computation and storage resources to the fullest, which is to pack the matrix $Z$ into a single ciphertext in a row-by-row manner:
\begin{align*}
  \begin{bmatrix}
    z_{[0][0]} & z_{[0][1]} & \ldots & z_{[0][f]} \\
    z_{[1][0]} & z_{[1][1]} & \ldots & z_{[1][f]} \\
    \vdots     & \vdots     & \ddots & \vdots     \\
    z_{[n][0]} & z_{[n][1]} & \ldots & z_{[n][f]}
  \end{bmatrix}
  \xmapsto{\text{ Database Encoding Method \cite{kim2018logistic} }}
  Enc\begin{bmatrix}
    z_{[0][0]} & z_{[0][1]} & \ldots & z_{[0][f]} \\
    z_{[1][0]} & z_{[1][1]} & \ldots & z_{[1][f]} \\
    \vdots     & \vdots     & \ddots & \vdots     \\
    z_{[n][0]} & z_{[n][1]} & \ldots & z_{[n][f]}
  \end{bmatrix}
\end{align*}
They treated the plaintext slots, a row vector of complex numbers, as a matrix $Z$ and developed various operations on the ciphertext to manage the data in the matrix $Z$, using mainly three HE basic operations: rotation, addition and multiplication.

This encoding scheme enables manipulation of the data matrix $Z$ by performing homomorphic encryption (HE) operations on the ciphertext $Enc[Z]$. Operations such as extracting specific columns or computing gradients in logistic regression models become achievable.

For instance, to filter out all columns except the first column of $Enc[Z]$, a constant matrix $F$ is designed with ones in the first column and zeros elsewhere. Multiplying $Enc[Z]$ by $Enc[F]$ yields the resulting ciphertext $Enc[Z_p]$:

\begin{align*}
  Enc[F] \otimes Enc[Z] &= Enc[Z_p] \quad (\text{where ``$\otimes$'' denotes component-wise HE multiplication}) \\
  &= Enc\begin{bmatrix}
    z_{[0][0]} & 0 & \ldots & 0 \\
    z_{[1][0]} & 0 & \ldots & 0 \\
    \vdots     & \vdots & \ddots & \vdots \\
    z_{[n][0]} & 0 & \ldots & 0
  \end{bmatrix}.
\end{align*}

Further operations, such as the ``$\texttt{SumColVec}$'' procedure introduced by Han et al. \cite{han2018efficient}, allow for more complex calculations on the ciphertexts.

In most real-world cases, the training dataset is too large to be encrypted into a single ciphertext. Han et al. \cite{han2018efficient} divided the dataset vertically into multiple fixed-size sub-matrices each of which encoded a fixed number (a power of two) of columns from the original matrix $Z$. If the last sub-matrix didn't have enough columns to be filled, zero columns would be padded in the end. Ciphertexts encoded this way can be bootstrapped efficiently but each ciphertext doesn't encrypt an intergrate example.

Since Kim et al.~\cite{kim2018logistic} worked on small dataset, their vertical partition is not be for the large dataset. However, their follow-up work~\cite{han2018efficient} solve this problem and divide $Z$ into multiple $m \times g$ sub-matrices $Z_{[i][j]}$ as follows:
\begin{align*}
Z_{[i][j]} = 
  \begin{bmatrix}
    z_{[0][0]} & z_{[0][1]} & \ldots & z_{[0][f]} \\
    z_{[1][0]} & z_{[1][1]} & \ldots & z_{[1][f]} \\
    \vdots     & \vdots     & \ddots & \vdots     \\
    z_{[n][0]} & z_{[n][1]} & \ldots & z_{[n][f]}
  \end{bmatrix}
\end{align*}
where $0 \le i \le n/m$ and $0 \le j \le f/g$. 

The number $m$ and $g$ are both set to a power of two such that their product is the number of maximum ciphertext slots, that is, $m \times g = N/2$. This will affect how to partition the weight vector in order to efficiently calculate the gradient in the training process, which in turn determains the size of a mini-batch block. Like the vertical partition~\cite{kim2018logistic}, zero rows and columns will be padded in some limb submatrices in order to make [it] a power of two. Each sub-matrix $Z_{[i][j]}$ is supposed to be packed into a single cipheretxt in a row by row manner. They first represent the sub-matrix in a vector and then encrypt this vector using the $\texttt{SIMD}$ technique. At the end of the day, they encrypt the whoel training dataset into  $ 
(n+1)(1+f)/mg$ ciphertexts.

\begin{figure}[htp]%[htp]
\centering
\includegraphics[width=5.5in]{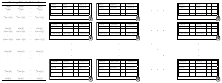} %[width=6in]
\caption{\protect\centering Partition and Encryption of Training Data }
\label{Dataset MiniBatch Partition ALL}
\end{figure}

\subsection{Logistic Regression}
Logistic regression (LR) is a widely-used machine learning algorithm with simple structures to learn a 2-layer neural network without any hidden layers for both binary and multiclass classification tasks. There are some works focusing the privacy-preserving multiclass logistic regression training based on mere HE, but in this work we only discuss the binary version.
In logistic regression for binary classification tasks, the goal is to determine whether a binary-valued variable belongs to a particular class by using the sigmoid function $\sigma(z)=1/(1+\exp(-z))$.  Supposing that a weight vector $w \in \mathbb{R}^{(1+f)}$, a vector $x_i$ of an input sample with $f$ features and its class label $y_i \in \{-1,+1\}$\footnote{The label can also take the form $y_i \in \{0,1\}$, resulting in a different gradient formulation.}, we force:
\begin{equation*}
  \begin{aligned}
    \Pr(y=+1|\mathbf{x}, \boldsymbol{w}) &= \sigma(\boldsymbol{w}^{\top} \mathbf{x}) = \frac{1}{1+e^{-\boldsymbol{w}^{\top} \mathbf{x}}},
  \end{aligned}
\end{equation*}
and hence
\begin{equation*}
  \begin{aligned}
    \Pr(y=-1|\mathbf{x}, \boldsymbol{w}) &= 1 - \Pr(y=+1|\mathbf{x}, \boldsymbol{w}) = \frac{1}{1+e^{\boldsymbol{w}^{\top} \mathbf{x}}}.
  \end{aligned}
\end{equation*}
Combining the above two equations, we obtain one unified equation:
\begin{equation*}
  \begin{aligned}
    \Pr(y=y_i|\mathbf{x}, \boldsymbol{w}) &= \frac{1}{1+e^{-y_i \boldsymbol{w}^{\top} \mathbf{x}}}.
  \end{aligned}
\end{equation*}
LR employs a threshold, typically $0.5$, compared with the output and determine the resulting class label.

It is commonly assumed that each sample in the dataset is independent and identically distributed (i.i.d.). Therefore, the probability of all samples in the dataset being true (occurring together) can be approximated by  
\[
L(\boldsymbol{w}) = \prod_{i=1}^{n} \Pr(y_i \mid \mathbf{x}_i, \boldsymbol{w}).
\] According to the maximum likilihood estimate, LR aims to find a parameter $\boldsymbol{w}$ to maximize $L(\boldsymbol{w})$, or its log-likelihood function $\ln L(\boldsymbol{\beta})$ for convient computation:\footnote{A more common way is to minimize the negative log likelihood function.}
\begin{equation*}
  \begin{aligned}
    l(\boldsymbol{w}) = \frac{1}{n} \ln L(\boldsymbol{w})= - \frac{1}{n}  \sum_{i=1}^{n} \ln(1+e^{-y_{i}\boldsymbol{w}^{\top} \mathbf{x}_i}),
  \end{aligned}
\end{equation*}
where $n$ is the number of samples in the training dataset ${(x_i, y_i)}_n$. The function $l(\boldsymbol{w})$ is convex but in some cases doesn't reach its maximization point~\cite{Allison2008LRConvergenceFail}. Fortunately, the vector \(\boldsymbol{w}\) that maximizes \(l(\boldsymbol{w})\) can still be obtained with high accuracy by employing iterative optimization techniques. These include gradient ascent algorithms and the Newton-R method, which utilize the gradient \(\nabla_{\boldsymbol{w}} l(\boldsymbol{w})\) and the Hessian matrix \(\nabla_{\boldsymbol{w}}^2 l(\boldsymbol{w})\) of the log-likelihood function as follows:
\begin{equation*}
  \begin{aligned}
    \nabla_{\boldsymbol{w}} l(\boldsymbol{w}) &= \frac{1}{n} \sum_i (1 - \sigma(y_i \boldsymbol{w}^{\top} \mathbf{x}_i))y_i \mathbf{x}_i, \\
    \nabla_{\boldsymbol{w}}^2 l(\boldsymbol{w}) &= \frac{1}{n} X^{\top}SX,
  \end{aligned}
\end{equation*}
where $S$ is a diagonal matrix with entries $S_{ii} = (\sigma(y_i \boldsymbol{w}^{\top} \mathbf{x}_i) - 1)\sigma(y_i \boldsymbol{w}^{\top} \mathbf{x}_i)$ and $X$ represents the dataset.

\textbf{Nesterov Accelerated Gradient Optimizer} Like~\cite{han2018efficient}, we choose Nesterov Accelerated Gradient (NAG) as the gradient descent optimization algorithm except using a faster gradient variant to accelerate it. Among various optimization methods, NAG is perhaps the best efficient candidate with decent optimization performance since other algorihtms such as Adagrad-like methods are not HE-friendly due to the freqcent expensive division operation. The gradient $ascent$ method for NAG can be formulated as follows:
\begin{align}
  w_{i+1} &=  \boldsymbol{v}_{i} + \gamma_i \cdot \nabla_w l(\boldsymbol{v}_i), \label{first formula} \\
  \boldsymbol{v}_{i+1} &= (1-\eta_i) \cdot  w_{i+1} + \eta_i \cdot  w_{i},  
\end{align}
where $w_i$ and $v_i$ are two weight vectors to update at each iteration $i$, and $\gamma_i$ and $\eta_i$ are NAG's control parameters.

\section{Technical Details}
In this section, we introduce a mini-batch version of the efficient logistic regression (LR) algorithm enhanced by a faster gradient variant called quadratic gradient~\cite{chiang2022privacy}. We first introduce the concept of quadratic gradient and then investigate potential mini-batch formulations of algorithms that exploit this gradient variant. We also discussed strategies for choosing appropriate learning rates for these mini-batch formulations. Finally, we show that the full-batch algorithm on large datasets can still be performed efficiently even in the encrypted setting, even though the dataset is encoded and encrypted in the same way as in the mini-batch approach.

\subsection{Quadratic Gradient}
In 2019, Chiang~\cite{chiang2022privacy} proposed an efficient and promising gradient variant, $\texttt{quadratic gradient}$, and claimed that it can accelerate various gradient descent/ascent algorithms with the curve information. Indeed, this gradient variant can combine first-order gradient methods with the Hessian matrix, resulting in various new algorithms with clear performance in convergenceing speed.
%The method known as Chiang's Quadratic Gradient (CQG)~\cite{,chiang2022quadratic,chiang2022privacy} presents a promising and expedited gradient variant that combines first-order gradient descent/ascent algorithms with the second-order Newton–Raphson method. This acceleration of the raw Newton–Raphson method with various gradient algorithms holds potential for constructing super-quadratic algorithms. 
This $\texttt{quadratic gradient}$ could be a game-changer to the Newton–Raphson method, showing new ways of using Hessian matrix and possiblely supertiding the line-searching technique widely-adopted currently. 

For a function $F(x)$ with its gradient $g$ and Hessian matrix $H$, the construction of $\texttt{quadratic gradient}$ begins by forming a diagonal matrix $\bar{B}$ from the Hessian $H$ itself:

\[
\bar B = 
\begin{bmatrix}
  \frac{1}{ \epsilon + \sum_{i=0}^{f} | \bar h_{0i} | } & 0 & \ldots & 0 \\
 0 & \frac{1}{ \epsilon + \sum_{i=0}^{f} | \bar h_{1i} | } & \ldots & 0 \\
 \vdots & \vdots & \ddots & \vdots \\
 0 & 0 & \ldots & \frac{1}{ \epsilon + \sum_{i=0}^{f} | \bar h_{fi} | } \\
\end{bmatrix},
\]
where $\bar h_{ji}$ represents the elements of  matrix $H$, and $\epsilon$ a small positive constant.

The $\texttt{quadratic gradient}$ for function $F(\mathbf x)$, defined as $G = \bar B \cdot g$, has the same dimension as the original gradient $g$. As a result, in practice, we can apply $G$ in a similar way as $g$, except that: (1) the naive gradient is replaced with the quadratic gradient; and (2) a new learning rate $\eta$ is adopted (usually increased by $1$). For instance, $\mathbf x = \mathbf x + \eta G$ can be used to maximize the function $F(\mathbf x)$ while $\mathbf x = \mathbf x - \eta G$ to minimize. Sophisticated gradient (descent) methods, such as NAG and Adagrad, can be applied  to further improve performance.

To efficiently apply $\texttt{quadratic gradient}$, it is advisable to attempt obtaining a good bound constant matrix to replace the Hessian itself even though such fixed-Hessian subsititcane might not exist for most functions. 
 
NAG with quadratic gradient replaces \eqref{first formula} with $w_{i+1} =  \boldsymbol{v}_{i} + (1 + \gamma_i) \cdot \bar B \cdot \nabla_w l(\boldsymbol{v}_i)$.

\subsection{Mini-Batch Method}
%mention the Han's Learning Rate Setting and maybe do some experiments about it

%https://www.kaggle.com/code/residentmario/full-batch-mini-batch-and-online-learning

In large-scale machine learning, the stochastic gradient descent (SGD) method, which processes one sample at a time, is inefficient and does not fully exploit the potential of encrypted data. The mini-batch gradient method addresses this issue by dividing the dataset into smaller batches. The training process consists of multiple epochs, with each epoch containing several iterations, where one mini-batch is processed per iteration. This approach improves computational efficiency, regularization, convergence speed, and memory usage.

The batch size, a hyperparameter in mini-batch algorithms, significantly impacts training. Smaller batch sizes introduce more noise but allow for faster updates, whereas larger batch sizes reduce noise, although they slow down training. In an encrypted setting, smaller batch sizes require more homomorphic operations and consequently more bootstrapping, making them less efficient. In clear-text training, a batch size of 32 or 64 is commonly used, but in encrypted contexts, batch sizes of 512 or 1024 are recommended. This choice is particularly beneficial for mini-batch algorithms utilizing the Hessian matrix, as larger batches capture more curve information, which enhances training performance.

The learning rate, another crucial hyperparameter, must be adapted when transitioning from full-batch to mini-batch methods. In the full-batch algorithm, the learning rate is typically decreased exponentially from a value greater than 1.0 to 1.0, ensuring convergence in later stages. For mini-batches, however, the learning rate must decrease below 1.0, approaching zero, because the Hessian matrix for each mini-batch differs from that of the entire dataset, potentially impeding convergence.

This paper proposes an exponential decay learning rate mechanism, denoted as \( F_2 \), for our mini-batch algorithm, designed to meet the following objectives: at \( t = 0 \), the learning rate should be at its maximum value, \( \text{max} \); at \( t = T \), it should reach the minimum value, \( \text{min} \); and the decay rate is controlled by the parameter \( \gamma \). The function \( F_2 \) is defined as:

\[
F_2(t) = \text{max} - (\text{max} - \text{min}) \cdot \left(\frac{t}{T}\right)^\gamma
\]

where \( \gamma \) is the parameter controlling the rate of decay. A larger value of \( \gamma \) results in a slower rate of decay, meaning the learning rate decreases more gradually.
In the function \( F_2(t) \), the variable \( t \) represents the current iteration count. The value \( \text{max} \) is the initial learning rate while the value \( \text{min} \) is the target learning rate at the final iteration.

%These function forms allow for precise control over the behavior of the learning rate throughout the iterative process, with the parameter \( \gamma \) providing flexible adjustment of the decay rate.

The resulting mini-batch algorihtm for the enhanced full-batch algorithm~\cite{chiang2022privacy} is shown in Algorithm~\ref{ alg:enhanced mini-batch nag's algorithm }. It is important to note that our mini-batch enhanced NAG approach requires computing the $\bar{B}_i$ for each mini-batch.

\begin{algorithm}[hp]
    \caption{Enhanced mini-batch NAG}
     \begin{algorithmic}[1]
        \Require training-data mini-batches $\{ X_i | X_i \in \mathbb{R} ^{m \times (1+f)}, i \in [1, m] \}$ where $m$ is the number of mini-batches. Namely, it will take $m$ iterations (passes) to complete one epoch; training-label mini-batches, $\{Y_i | Y_i \in \mathbb{R}^m , i \in [1, m] \}$, for each corresponding training-data mini-batches $X_i$ ; the number of epoches,  $E$; and the number of iterations,  $K$;
        \Ensure the parameter vector $ W \in \mathbb{R} ^{(1+f)} $ 
        
        \For{$i := 1$ to $m$}
           \State  Compute $\bar B$ from $X_i$ using the full-batch enhanced NAG algorihtm, resulting in $\bar B_i$ for each mini-batch 
           \Comment{$\bar B_i \in \mathbb R^{(1+f) \times (1+f)}$}
        \EndFor
        
        \State Set $ W \gets \boldsymbol 0$
        \Comment{Initialize the weight vector $\boldsymbol{W} \in \mathbb{R} ^{(1+f)} $}
        \State Set $ V \gets \boldsymbol 0$
        \Comment{Initialize the vector $\boldsymbol{V} \in \mathbb{R} ^{(1+f)} $}
       
        \State Set $\alpha_0 \gets 0.01$
        %\Comment{Initialize the weight vector $\boldsymbol{v} \in \mathbb{R} ^{(1+d)} $}
        \State Set $\alpha_1 \gets 0.5 \times (1 + \sqrt{1 + 4 \times \alpha_0^2} )$ 
        %\Comment{Initialize the weight vector $\boldsymbol{v} \in \mathbb{R} ^{(1+d)} $}
        
        \For{$epoch := 0$ to $E - 1$}
           \State Shuffle the mini-batches $\{X_i\}$ or Not
           %\State (or just shuffle the indexes pointing to mini-batches for efficient programming)
           
           \For{$k := 1$ to $m$}
              %\LeftLineCommentCont{The Iterative Procedure of Gradient Descent Method}
              \State Set $Z \gets \boldsymbol 0 $
              \Comment{$Z \in \mathbb{R}^{n}$ is the inputs to sigmoid function}
              \For{$i := 1$ to $n$}
                 \For{$j := 0$ to $f$}
                    \State $ Z[i] \gets  Z[i] +  Y_k[i] \times  W[j] \times  X_k[i][j] $
                    \Comment{$X_k[i][j]$}
                 \EndFor
              \EndFor
              %\LineCommentCont{To compute the value of the sigmoid function for each input $ Z_{i} $}
              %\LineCommentCont{To compute the value of the sigmoid function for each input $ Z_{i} $ : $\boldsymbol \sigma \in \mathbb{R}^{n}$}
              \State Set $\boldsymbol \sigma \gets \boldsymbol 0 $
              %\Comment{$\boldsymbol \sigma \in \mathbb{R}^{n}$ is to store the outputs of the sigmoid function}
              \Comment{$\boldsymbol \sigma \in \mathbb{R}^{n}$}
              \For{$i := 1$ to $n$}
                 \State $\boldsymbol \sigma[i] \gets 1 / (1 + \exp (-Z[i])) $
              \EndFor
              % # g = [Y@(1 - sigm(yWTx))]T * X
              %\LineCommentCont{To calculate the gradient $\boldsymbol g \in \mathbb{R}^{(1+d)} $}
              \State Set $\boldsymbol g \gets \boldsymbol 0$
              \For{$j := 0$ to $f$}
                 \For{$i := 1$ to $n$}
                    \State $\boldsymbol g[j] \gets \boldsymbol g[j] + (1 - \boldsymbol \sigma[i] ) \times  Y_k[i] \times X_k[i][j] $
                 \EndFor
              \EndFor
              %\LineComment{To calculate the quadratic gradient $ G \in \mathbb{R}^{(1+d)}$ }
              \State Set $ G \gets \boldsymbol 0$
              \For{$j := 0$ to $f$}
                 \State $ G[j] \gets \bar B[j][j] \times \boldsymbol g[j]$
              \EndFor
              %\LineComment{To update the weight vector $V$ }
           	  %eta = (1 - alpha0) / alpha1
	          %gamma = 1.0/(iter+1)/MX.shape[0]
              \State Set $\eta \gets (1 - \alpha_0) / \alpha_1$
              \State Set $\gamma \gets  1 / n \times F_2(t) $  
              \Comment{$F_2(t)$ is the adopted learning rate setting. }
              
              \For{$j := 0$ to $f$}
                 \State $ V_{temp} \gets W[j] + (1 + \gamma)  \times G[j] $
                 \State $ W[j] \gets (1 - \eta) \times V_{temp} + \eta \times V[j] $
                 \State $ V[j] \gets V_{temp} $
              \EndFor
              \If {$epoch \times E + k \ge K $}
                 \State \Return $ W $
              \EndIf
              %alpha0 = alpha1
	          %alpha1 = (1. + sqrt(1. + 4.0 * alpha0 * alpha0)) / 2.0
	          \State $\alpha_0 \gets \alpha_1$
              \State $\alpha_1 \gets 0.5 \times (1 + \sqrt{1 + 4 \times \alpha_0^2} )$ 
           \EndFor
        \EndFor
        \State \Return $ W $
        \end{algorithmic}
       \label{ alg:enhanced mini-batch nag's algorithm }
\end{algorithm}

\paragraph{Performance Evaluation} We evaluate our enhanced mini-batch NAG algorihtm in the clear state using Python on two datasets from~\cite{han2018efficient} and compare these results with the baseline algorithm~\cite{han2018efficient} in terms of convergen speed using only the loss function (log-likelihood function) as the indicator. The batch size is set to the same as in~\cite{han2018efficient}: 512 for the financial dataset and  1024 for the restructured MNIST dataset. They set the learning rate for their first-order mini-batch algorihtm to $\frac{0.01}{n}$ for the Financial dataset and $\frac{1.0}{n}$ for the restructured MNIST dataset while we set $1 + \frac{0.01}{n}$ for the Financial dataset and $1 + \frac{1.0}{n}$ for the restructured MNIST dataset. Figure~\ref{fig0} shows that our mini-batch enhanced method converges faster than the first-order baseline algorithm on both training sets of the datasets.

\begin{figure}[ht]
\centering
\captionsetup[subfigure]{justification=centering}
%\quad
%%%%%%%%%%%%%%%%%%%%%%%%%%%%%%%%%%%%%%%%%%%%%%
\subfloat[The MNIST  dataset]{%
\begin{tikzpicture}%[scale=.7]
\scriptsize  % control the font size of xylabels and Title
\begin{axis}[
    width=7cm, % the width of x-axis
    %width=0.6\textwidth,
    %height=0.6\textwidth,
    %title={Temperature dependence of solubility},
    xlabel={Iteration Number},
    ylabel={Maximum Likelihood Estimation},
    y label style={at={(+0.04, 0.5)}},
    xmin=0, xmax=30,
    legend pos=south east,
    legend style={nodes={scale=0.7, transform shape}},
    legend cell align={left},
    xmajorgrids=true,
    ymajorgrids=true,
    grid style=dashed,
]
\addplot[
    color=black,
    mark=triangle,
    mark size=1.2pt,
    ] 
    table [x=Iterations, y=NAG, col sep=comma] {PythonExperimentforSection32NAGvsNAGGMNISTMLE.csv};
\addplot[
    color=blue,
    mark=diamond*, 
    mark size=1.2pt,
    densely dashed
    ]  
    table [x=Iterations, y=NAGG, col sep=comma] {PythonExperimentforSection32NAGvsNAGGMNISTMLE.csv};
   \addlegendentry{NAG}
   \addlegendentry{Enhanced NAG}
\end{axis}
\end{tikzpicture}
\label{fig:subfig01}}
%%%%%%%%%%%%%%%%%%%%%%%%%%%%%%%%%%%%%%%%%%%%%%
\subfloat[The Financial dataset]{%
\begin{tikzpicture}%[scale=.7]
\scriptsize 
\begin{axis}[
    width=7cm, 
    xlabel={Iteration Number},
    xmin=0, xmax=30,
    legend pos=south east,
    legend style={yshift=0.5cm,nodes={scale=0.7, transform shape}},
    legend cell align={left},
    %legend pos = outer north east,
    xmajorgrids=true,
    ymajorgrids=true,
    grid style=dashed,
]
\addplot[
    color=black,
    mark=triangle,
    mark size=1.2pt,
    ] 
    table [x=Iterations, y=NAG, col sep=comma] {PythonExperimentforSection32NAGvsNAGGCreditMLE.csv};
\addplot[
    color=blue,
    mark=diamond*, 
    mark size=1.2pt,
    densely dashed
    ]  
    table [x=Iterations, y=NAGG, col sep=comma] {PythonExperimentforSection32NAGvsNAGGCreditMLE.csv};
   \addlegendentry{NAG}
   \addlegendentry{Enhanced NAG}
\end{axis}
\end{tikzpicture}
\label{fig:subfig02}}
%\quad
\caption{\protect\centering Training results in the unencrypted setting for the MNIST and Financial datasets}
%Main \subref{fig:subfig1} figure \ref{fig0} caption}
\label{fig0}
\end{figure}

\subsection{Full-Batch Method}
One limitation for our enhanced mini-batch algorithm is that like the first-order mini-batch algorithm the parameters of the learning-rate scheculer have to be tuned, which will to some extend damage the non-interactiaion property of HE. On the other hand, full-batch quadratic gradient algorithms don't have such annoying issues: we can just set the learning rate to $1$ and the algorihtm will still converge at a decent speed. As we discussed before, full-batch gradient methods is not suitable to apply on large datasets. However, in the encrypted domain, this is no longer a problem any more: each ciphertext can usually encrypt a large number of samples and compared with the time to refreshing ciphertexts homomorphic evaluation of updating the weight ciphertexts for the full-batch data might require less time and will not consume more modulus level. As a result, the full-batch gradient method using quadratic gradient might be more approprate than the mini-batch enhanced algorithm using quadratic gradient.

And it is possible and easy to perform the full-batch gradient method even though the dataset is divided into mini-batches and encrypted into multiple cipheretxts.

Moreover, the $\bar B$ in quadratic gradient $G = \bar B g$ for the whole dataset can be obtained from those of the mini-batches $\bar B_i$: $$\bar B[i][i] = \frac{1}{ \sum_{k=1}^n \frac{1}{\bar B_k[i][i]}} .$$

\section{ Secure Training }
In this section, we explain how to train the logistic regression model in the encrypted state using our mini-batch enhanced NAG method, the process of which is very similiar to that of the full-batch enhanced NAG method. 

First of all, an essential step of privacy-preserving logistic regression training is to evaluate the activation function, namely the sigmoid function $\sigma(x) = 1/(1+e^{-x})$. Since HE evaluator cannot compute non-polynomial functions, most of which, forunately, can be approximate by polynomials to arbitate precious. As a result, the common solution for HE to non-polynomials is to replace them with their corresponding  high precision polynomial approximations  over some intervals within a given fixed maximum error.

\subsection{Polynomial Approximation}
There are various known methods that can be used to find a polynomial approximation of (activation) functions. Taylor expansion and Lagrange interpolation are classic methods to find a polynomial approximation and provide a precise approximation but only on a small range to the point of interest with drastically increase of approximation error outside the given small range. On the other hand, the least squares approach aims to find a global polynomial approximation and provides a good approximation on a large range. It is so widely adopted in real-world applications, such as in~\cite{kim2018logistic, han2018efficient, chiang2022privacy}, that Python and Matlab have a function named `` $\texttt{polyfit($\cdot$)}$ '' to approximate non-polynomials in a least-squares sense. In addition, the minimax approximation algorithm trys to find a polynomial approximation that minimizes the maximus error of the given target function and ensures a good quality of a polynomial approximation at each point of the approximation interval.

There are also researches on polynomial approximation on large intervals. Cheon et al.~\cite{cheon2022homomorphicevaluation} introduce domain extension polynomials that can be repeatedly used with to perform iterative domain-extension precess, efficiently approximating sigmoid-like functions over significantly large intervals. This efficient solution for homomorphic evaluation on large interals cannot deal with all shape functions.

For a fair comparision with~\cite{han2018efficient}, we adopt their degree $3$ polynomial approximations using the least square approach to approximate the sigmoid function over the domain intervals $[-8, +8]$ and $[-16, +16]$, respectively: 
$$ g_{8}(x) = 0.5 + 0.15 \cdot x  - 0.0015 \cdot x^3, $$ 
and
$$ g_{16}(x) = 0.5 + 0.0843 \cdot x  - 0.0002 \cdot x^3 . $$

\subsection{Usage Model}
Like~\cite{han2018efficient}, our approach can be used in several usage scenarios. 

One typical scenario is in which there is only one party that own the private data. This party first encrypts its sensitive data with some given HE libiray, uploads the resulting ciphertexts to a cloud server and later downloads the target encrypted model from the same cloud that has performed the machine learining training on the uploaded and encrypted data. The single party exclusively owns the private key of the HE scheme and and don't need to share the secret key with anyone else while the cloud can have various public parameters and evaluation public keys.

Another situation in which our approach can also be helpful is the mutiple party owning their different data. In that case, the public key of the HE scheme is issued to each data owner and all the data is encrypted with the same public key under the same HE scheme. The cloud does the same task as before and outputs an encrytped model which can be decrypted by a single entity who owns the private key, or a group of entities having a random share of the private key~\cite{jain2017threshold,cramer2001multiparty}, jointly generating the public key by an additional key sharing protocol.

Note that, as long as the underlying HE scheme works normalily and its secret key is not disclosed, no infromation other than the encrypted well-learned model is revealed to each other in the protocol, even if the cloud is compromised. Hence, our approach, as well as others based on mere HE shceme could be the ultimate solution for privacy-preserving machine learning applications on analyzing sensitive data while keeping privacy.

\subsection{Whole Pipeline}
Our approach can be deployed in several practical scenarios, which could be the model in \cite{han2018efficient} or \cite{kim2018secure}. Without paying attention to the usage model, we assume that a third-party authority, which could even be the client itself, is needed to build a homomorphic encryption system. This third party issues secret key, public key, encryption algorithm, and decryption algorithm to the client. Another duty of this third party is to deploy the main HE system, including all public keys to the cloud, as well as the privacy-preserving LR training algorithm.

Given the training dataset $ X \in \mathbb{R}^{(1+n) \times (1+f)} $ and training label $ Y \in \mathbb{R}^{(1+n) \times 1} $, we adopt the same database-encoding method that Han et al.~\cite{han2018efficient} developed and for simplicity we assume that the whole dataset can be encrypted into a single ciphretext.[For the sake of  simplicity, we adopt the assumptions that the entire dataset can be encrypted into a single ciphertext, and that the training data have been normalized into the range $(0, 1)$ and the training labels have been transferred to $\{-1,+1\}$, if not. ] The client first combines $X$ with $Y$ into the matrix $Z$ and prepares  $\bar B$ using $ -\frac{1}{4} X^{\top} X $ for building quadratic gradient and restruct $\bar B$ into a vector form with its diagnoal elements. Next, the client encrypts the matrix $Z$, the vector form $\bar B$ and the weight vector $W^{(0)}$ into ciphertexts $ \text{ct}_{Z} $, $ \text{ct}_{\bar{B}} $ and $ \text{ct}_{W^{(0)}} $. The vector $\bar B$ and the weight vector are both copied $(1+n)$ times to fill the plaintext slots. Using the database encoding methods~\cite{kim2018logistic} , the resulting ciphertexts are described as follows:  
\begin{align*}
 \text{ct}_Z &= \text{Enc}
\begin{bmatrix}
 y_{[0]}  &   y_{[0]} x_{[1][1]} &  \ldots  &  y_{[0]} x_{[0][f]} \\
 y_{[1]}  &   y_{[1]} x_{[1][1]} &  \ldots  &  y_{[1]} x_{[1][f]} \\
 \vdots &   \vdots        &  \ddots  &  \vdots \\
 y_{[n]}  &   y_{[n]} x_{[n][1]} &  \ldots  &  y_{[n]} x_{[n][f]}
\end{bmatrix} =
\begin{bmatrix}
 z_{[0][0]}  &   z_{[0][1]} &  \ldots  &  z_{[0][f]} \\
 z_{[1][0]}  &   z_{[1][1]} &  \ldots  &  z_{[1][f]} \\
 \vdots &   \vdots        &  \ddots  &  \vdots \\
 z_{[n][0]}  &   z_{[n][1]} &  \ldots  &  z_{[n][f]}
\end{bmatrix} , \\
 \text{ct}_{\bar{B}} &= \text{Enc}
\begin{bmatrix}
 \bar{B}_{[0][0]} & \bar{B}_{[1][1]} &  \ldots  & \bar{B}_{[f][f]} \\
 \bar{B}_{[0][0]} & \bar{B}_{[1][1]} &  \ldots  & \bar{B}_{[f][f]} \\
 \vdots & \vdots & \ddots  & \vdots \\
 \bar{B}_{[0][0]} & \bar{B}_{[1][1]} &  \ldots  & \bar{B}_{[f][f]}
\end{bmatrix},
 \text{ct}_{W}^{(0)} = \text{Enc}
\begin{bmatrix}
 w_{0}^{(0)}  &   w_{1}^{(0)} &  \ldots  &  w_{f}^{(0)} \\
 w_{0}^{(0)}  &   w_{1}^{(0)} &  \ldots  &  w_{f}^{(0)} \\
 \vdots &   \vdots        &  \ddots  &  \vdots \\
 w_{0}^{(0)}  &   w_{1}^{(0)} &  \ldots  &  w_{f}^{(0)}
\end{bmatrix}, 
\end{align*}
Finally, the client sends these ciphertexts to the public cloud. The public cloud takes these ciphertexts $ \text{ct}_Z $, $ \text{ct}_{W}^{(0)} $ and $ \text{ct}_{\bar{B}} $ to evaluate the enhanced mini-batch NAG algorithm to find a decent weight vector. At each iteration, the algorithm evaluates the quadratic gradient and aims to update the modeling vector $W^{(t)}$. At the end of the training, the public server return to the client the ciphertext encrypting the resulting weight vector.

\begin{figure}[htp]%[htp]
\centering
\includegraphics[width=5.6in]{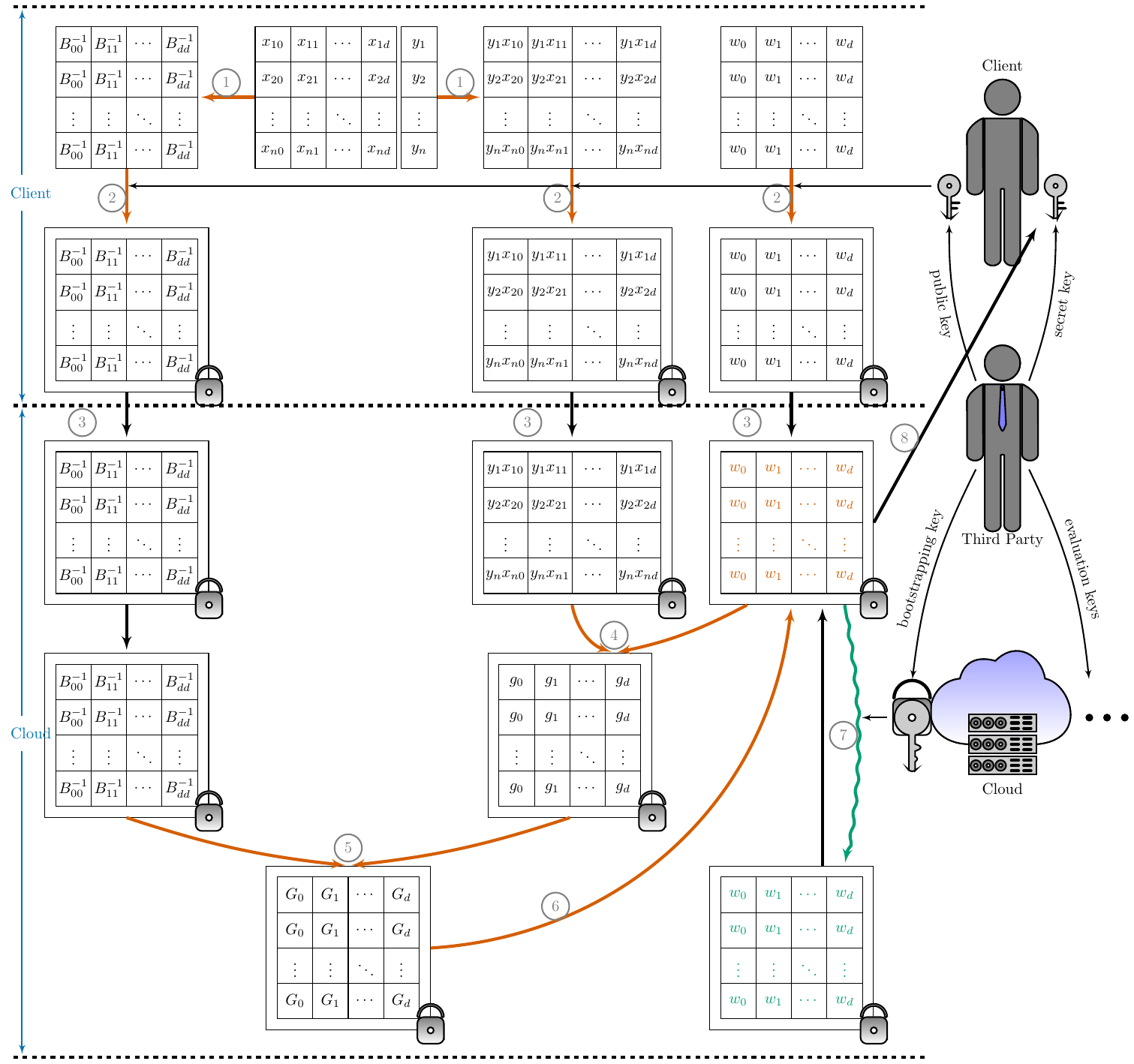} %[width=6in]
\caption{\protect\centering the entire process of logistic regression training via homomorphic encryption}
\label{Main Flow Chart}
\end{figure}

The full pipeline of privacy-preserving logistic regression training is illustrated in Figure~\ref{Main Flow Chart}, which consists of the following steps: For simplicity, we assume that the dataset is encrypted into a single ciphertext.

\indent $\texttt{ Step 1:}$ 
The client prepares two data matrices, $\bar B$ and $Z$, using the training dataset.

\indent $\texttt{ Step 2:}$ 
The client then encrypts three matrices, $\bar B$, $Z$ and weight matrix $W^{(0)}$, into three ciphertexts $ \text{ct}_Z $, $ \text{ct}_{W}^{(0)} $ and $ \text{ct}_{\bar{B}} $, using the public key given by the third party under the assigned HE system.   We recommend adopting the zero matrix as the initial weight matrix, if a already-trained weight matrix is not provided.

\indent $\texttt{ Step 3:}$
The client finally uploads the ciphertexts to the cloud server and finished its part of precess, waiting for the result.

\indent $\texttt{ Step 4:}$
The public cloud begins its work by evaluating the gradient ciphertext $ \text{ct}_{g}^{(0)} $ with $ \text{ct}_Z $ and $ \text{ct}_{W}^{(0)} $ using various homomorphic operations. Kim et al.~\cite{kim2018logistic} give a full and detail description about homomorphic evaluation of the gradient descent method.

\indent $\texttt{ Step 5:}$
The public cloud computes the quadratic gradient with one homomorphic multiplication from $ \text{ct}_{\bar{B}} $ and $ \text{ct}_{g}^{(0)} $, resulting in the ciphertext $ \text{ct}_{G}^{(0)} $.

\indent $\texttt{ Step 6:}$
This step is to update the weight ciphertext with the quadratic gradient ciphertext using our enhanced mini-batch NAG method. This will consume some modulus level of the weight ciphertext, finally .

\indent $\texttt{ Step 7:}$
This step checks if the remaining modulus level of the weight ciphertext enable another round of updating the weight ciphertext. If not, the algorithm would bootstrap only the weight ciphertexts using some public keys,  obtaining a new ciphertext encrypting the same weight but with a large modulus. 

\indent $\texttt{ Step 8:}$
This public cloud completes the whole iterations of homomorphic LR training, obtain the resulting weight ciphertext, and finishes its work by returning a ciphertext encrypting the updated modeling vector to the client. 

Now, the client could decipher the received ciphertext using the secret key and has the LR model for its only use.

%The full pipeline of privacy-preserving logistic regression training is illustrated in Figure~\ref{Main Flow Chart}, which consists of the following steps:

Han et al.~\cite{han2018efficient} give a detailed description of their HE-friendly LR algorithm with HE-optimized body of the iteration loop using HE programming in the encrypted domain. Our enhanced mini-batch NAG method is significantly similar to theirs except for needing one more ciphertext multiplication between the gradient ciphertext and the uploaded ciphertext encrypting $\bar B_i$ for each mini batch, and therefore please refer to~\cite{han2018efficient} for more information.

\section{Experiments}
We evaluate our enhanced mini-batch and full-batch algorithms of logistic regression training on two encrypted datasets from the baseline work~\cite{han2018efficient}: a real financial training dataset and the restructed MNIST dataset.  The C++ source code with the $\texttt{HEAAN}$ library is openly accessible at \href{https://github.com/petitioner/HE.LR}{$\texttt{https://github.com/petitioner/HE.LR}$}. 

For a fair comparison with~\cite{han2018efficient}, we use similar HE scheme setting, involving the following parameters: the regular scaling factor $\Delta$ (like for multiplications between ciphertexts); the constant scaling factor $\Delta_c$ used for multiplying constant matrices and scalars; the number of ciphertext slots $N/2$ and the initial ciphertext modulus $Q$. %Also, let $\texttt{wBits} = \log_2 \Delta$ and $\texttt{pBits} = \log_2 \Delta_c.$ 
We set these HE scheme parameters similarly to the baseline work~\cite{han2018efficient}. % while due to the difference of algorithms we set our learning rate to be $1$ greater than theirs.
For programming convenience, we adopted a unified parameter standard for 
N, Q, and $\Delta_c$ in both experiments presented in this paper. The differences in 
N and Q from the baseline result in variations in ciphertext size.  We selected a different constant scaling factor (for cMult) because our enhanced algorithm requires higher precision for $\bar B$. We also have the same number of ciphertext slots $N/2 = 2^{15}$ as the baseline work. The initial ciphertext modulus Q for the weight vectors 
W and V was selected through a trial-and-error process to ensure that one ciphertext refresh operation is needed per iteration. We selected 
Q such that one ciphertext refresh operation is required per iteration, thereby simplifying parameter selection. In contrast, the baseline work chooses different parameters for different experimental datasets. All experiments in the encrypted state in this section were executed on a public cloud with 8 vCPUs and 64 GB RAM, while the baseline used a machine with an IBM POWER8 (8 cores, 4.0 GHz) and 256 GB RAM. To simplify parameter settings, we used a consistent mini-batch size of 1024 across all experiments. In contrast, the baseline work uses different sizes for the two datasets: 512 for the Financial Dataset and 1024 for the MNIST Dataset.

\subsection{Financial Dataset}
We performed experiments on a real financial dataset to assess the efficiency and scalability of our algorithms. The encrypted dataset used to evaluate our logistic regression algorithm consists of real consumer credit information maintained by a credit reporting agency. It includes data from 1,266,325 individuals, each with 200 features relevant to credit rating, such as loan details, credit card usage, and delinquency records. The samples are labeled with a binary classification indicating whether the individual’s credit rating falls below a specified threshold.

\paragraph{Experimental Results}
Table 1 presents the detailed results of our experiment. We set the learning rate of both algorithms to be \( 2 - ( t/81 )^{2.5} \). The Mini-Batch algorithm performed 41 iterations, with each iteration taking approximately 45 seconds. In contrast, the Full-Batch algorithm required 57 iterations, with each iteration taking around 90 seconds. The Full-Batch algorithm has a higher average running time compared to the Mini-Batch algorithm because, in each iteration, Full-Batch processes all mini-batches, whereas the Mini-Batch algorithm processes only one mini-batch per iteration.

Both our Mini-Batch and Full-Batch algorithms obtained 79.5\% accuracy. This discrepancy can be attributed to the variation in precision of the encrypted model and the use of a baseline Polynomial Approximation, which does not perform well within its constrained range. Our enhanced algorithm, similar to the fixed Hessian method, could only achieve an AUROC of around 0.7, whereas the first-order gradient algorithm achieved an AUROC of 0.8.

We remark that both of our algorithms require fewer than 60 iterations to achieve optimal performance, whereas the baseline method needed 200 iterations to converge.

\begin{table}[h!]
\centering
\caption{Result of machine learning on encrypted data}
\label{tab2}
\begin{tabular}{|l|c|c|c|}
\hline
\multicolumn{1}{|l|}{Datasets}  & \multicolumn{3}{|c|}{Financial}  \\
\hline
\multicolumn{1}{|l|}{No. Samples (training)}  & \multicolumn{3}{|c|}{422,108}   \\
\hline
\multicolumn{1}{|l|}{No. Samples (validation)}  & \multicolumn{3}{|c|}{844,217}  \\
\hline
\multicolumn{1}{|l|}{No. Features}  & \multicolumn{3}{|c|}{200}  \\
\hline
Methods & \cite{han2018efficient}  &  Mini-Batch  & Full-Batch \\
\hline
\multirow{1}{*}{No. Iterations} & \multirow{1}{*}{200} & \multirow{1}{*}{41} & \multirow{1}{*}{57} \\
\hline
Learning Rate & 0.01 & $2 - (t/81)^{2.5}$ & $2 - (t/81)^{2.5}$  \\
\hline
Block Size (mini-batch) & 512 & 1,024 & 1,024 \\
\hline
Accuracy & 80\% & 79.5\% & 79.5\% \\
\hline
AUROC & 0.8 & 0.687 & 0.686  \\
\hline
Encrypted Block Size & 4.87 MB & 4.6 MB & 4.6 MB  \\
\hline
Average  Weight-Updating Time & --- &  $\approx$ 5 sec  & $\approx$ 60 sec  \\
\hline
Average  Weight-Refreshing Time & --- &  $\approx$ 38 sec  & $\approx$ 30 sec  \\
\hline
\# of iterations per bootstrapping  & 5 & 1  & 1  \\
\hline
Average Running Time & 318 sec & $\approx$ 45 sec  & $\approx$ 90 sec  \\
\hline
Ciphertexts $ \text{ct}_{\bar{B}}$ Size  &   ---   &  $27 \times 4.6 = 124.2$ MB   &  $1 \times 4.6 = 4.6$ MB   \\
\hline
HE Scheme Parameter: N & $2^{16}$  & $2^{16}$ & $2^{16}$  \\
\hline
HE Scheme Parameter: Q & $2^{635}$ & $2^{275}$ & $2^{275}$  \\
\hline
HE Scheme Parameter: $\Delta$ & $2^{30}$ & $2^{30}$ & $2^{30}$  \\
\hline
HE Scheme Parameter: $\Delta_c$ & $2^{15}$ & $2^{20}$ & $2^{20}$  \\
\hline
Polynomial Approximation & $g_{8}$ & $g_{8}$ & $g_{8}$  \\
\hline
\end{tabular}
\end{table}

%\paragraph{Microbenchmarks}

\subsection{MNIST Dataset}
We performed our logistic regression algorithms on the public MNIST dataset to conduct a more comprehensive evaluation.
We took the MNIST dataset restructured by the baseline, resulting in a binary classification problem between digits 3 and 8. The original images of 28×28 pixels were compressed to 14×14 pixels by averaging each 2×2 pixel block. The restructured dataset contains 11,982 training samples and 1,984 validation samples.

\paragraph{Experimental Results}
Table 2 summarizes the experimental results. Our mini-batch logistic algorithm required 32 iterations to train an encrypted model, taking 132 minutes in total, with an average iteration time of approximately 4 minutes. Similarly, the full-batch logistic algorithm took 145 minutes to complete 36 iterations, averaging around 6 minutes per iteration. These runtime characteristics are consistent with those observed on the financial dataset. After decrypting the learned models, we evaluated them on the validation dataset, achieving accuracies of 96.4\% and 96.9\%, respectively.

\begin{table}[htbp]
\centering
\caption{Result of machine learning on encrypted data}
\label{tab1}
\begin{tabular}{|l|c|c|c|c|c|c|}
\hline
\multicolumn{1}{|l|}{Datasets}    & \multicolumn{3}{|c|}{MNIST} \\
\hline
\multicolumn{1}{|l|}{No. Samples (training)}    & \multicolumn{3}{|c|}{11,982} \\
\hline
\multicolumn{1}{|l|}{No. Samples (validation)}   & \multicolumn{3}{|c|}{1,984} \\
\hline
\multicolumn{1}{|l|}{No. Features}    & \multicolumn{3}{|c|}{196} \\
\hline
Methods  & \cite{han2018efficient} & Mini-Batch & Full-Batch \\
\hline
\multirow{1}{*}{No. Iterations} &  \multirow{1}{*}{32} & \multirow{1}{*}{25} & \multirow{1}{*}{26} \\
\hline
Learning Rate &  1.0 & $2 - (t/36)^{2.5}$ & $2 - (t/36)^{2.5}$ \\
\hline
Block Size (mini-batch)  & 1,024 & 1,024 & 1,024 \\
\hline
Accuracy & 96.4\% & 96.33\% & 96.27\% \\
\hline
AUROC & 0.99 & 0.992 & 0.99 \\
\hline
Encrypted Block Size & 3.96 MB &  4.6 MB &  4.6 MB \\
\hline
Average  Weight-Updating Time &  ---  & $\approx$ 12 sec & $\approx$ 60 sec \\
\hline
Average  Weight-Refreshing Time &  ---  & $\approx$ 130 sec & $\approx$ 126 sec \\
\hline
\# of iterations per bootstrapping  & 3 & 1  & 1  \\
\hline
Average Running Time & 247 sec & $\approx$ 142 sec & $\approx$ 186 sec \\
\hline
Ciphertexts $ \text{ct}_{\bar{B}}$ Size &  ---  &  $84 \times 4.6 = 386.4$ MB  &  $7 \times 4.6 = 32.2$ MB  \\
\hline
HE Scheme Parameter: N & $2^{16}$ &$2^{16}$ & $2^{16}$  \\
\hline
HE Scheme Parameter: Q & $2^{495}$ & $2^{275}$ & $2^{275}$  \\
\hline
HE Scheme Parameter: $\Delta$ & $2^{40}$ & $2^{30}$ & $2^{30}$  \\
\hline
HE Scheme Parameter: $\Delta_c$ & $2^{15}$  & $2^{20}$ & $2^{20}$  \\
\hline
Polynomial Approximation & $g_{16}$  & $g_{16}$ & $g_{16}$  \\
\hline
\end{tabular}
\end{table}

\subsection{Discussion}
Based on the above experiments, the following conclusions can be drawn: Both our mini-batch and full-batch algorithms, enhanced with quadratic gradients, demonstrate faster convergence compared to traditional first-order gradient methods. Although the full-batch algorithm requires more time for weight updates due to the need to process all mini-batches, the additional time is negligible when compared to the time required for ciphertext refreshing. The mini-batch algorithm, processing one mini-batch at a time, results in shorter weight update times. However, it requires the pre-computation of $\bar{B} $  for each mini-batch, which increases the ciphertext transmission from the data owner to the cloud server. Furthermore, the mini-batch algorithm demands more careful selection of the learning rate to ensure convergence, while the full-batch algorithm can simply use a fixed learning rate of 1 to avoid issues with learning rate tuning. Finally, the full-batch algorithm generally requires higher precision when encrypting \(\bar{B}\), as the \(\bar{B}\) for the entire dataset is typically smaller than that of a mini-batch.

\iffalse
\paragraph{Limitations and Potential Solutions} The primary constraints of both the NAG method and the Enhanced NAG method in a privacy-preserving context involve the necessity to update not only the weight ciphertext $\boldsymbol{\beta}{t}$ but also the ciphertext encrypting $V{t+1}$. While this poses no significant issue in LR training, it becomes a time-consuming task in neural network training, where the weight is a matrix. Moreover, if we continue to use either the NAG method or the enhanced method, refreshing the $V_{t+1}$ ciphertext is also required. A potential solution could be the development of algorithms grounded in the raw quadratic gradient method itself.
\fi

\section{Conclusion}
In this paper, we presented an efficient algorithm for privacy-preserving logistic regression training on large datasets, and evaluated it over both the private financial data and the public restructed MNIST dataset.

\bibliography{HE.LRtraining}
\bibliographystyle{apalike}

\end{document}